# PECR: A Reproducible Specification and Synthetic Stress Test of Telemetry-Informed Vulnerability Prioritization for SD-WAN

PECR for SD-WAN Vulnerability Prioritization

SAEED ALAM

School of Computing Sciences and Computer Engineering, University of Southern Mississippi, Hattiesburg, MS, USA

saeed.alam@usm.edu | ORCID 0009-0004-8089-5658

Software-defined wide-area networking concentrates operational authority in controllers, orchestrators, and Internet-facing edges, but severity-only remediation queues do not represent current exposure or evidence quality. This paper specifies Predictive Exposure and Cryptographic Readiness (PECR), a decision-support method that combines nine normalized severity, threat, reachability, and consequence factors while reporting confidence and score intervals separately. Cryptographic dependencies enter an independent migration queue. The evaluation uses 100 role- and zone-conditioned synthetic vulnerability–asset records over 62 role-consistent assets, plus 30 independent generator replications. Against the operational PECR order, CVSS-only, EPSS-only, KEV-first, and CVSS×EPSS queues yield Kendall τ values of 0.196, 0.221, 0.301, and 0.234; a stronger five-factor context-lite comparator yields 0.755. Across 30,000 joint Dirichlet weight draws, median τ is 0.836, 0.893, and 0.924 under broad, moderate, and narrow perturbations. Masking 10% of factor cells labels 47 records as evidence-limited because their intervals cross a decision boundary. The study establishes executable specification, synthetic queue differentiation, and stress-test robustness. It does not establish predictive accuracy, avoided loss, or production effectiveness because no real outcome labels are used.

**CCS CONCEPTS • Security and privacy → Network security; Vulnerability management; Systems security • Networks → Network manageability**

Additional Keywords and Phrases: vulnerability prioritization, SD-WAN, attack paths, evidence quality, global sensitivity analysis, post-quantum cryptography, cryptographic bill of materials

## 1 INTRODUCTION

Software-defined wide-area networking (SD-WAN) replaces fixed private-WAN topologies with encrypted overlays, centralized policy, and multiple underlay links. The model improves path selection and service agility, but it also concentrates authority. An orchestrator can provision every branch, controllers distribute routing and security policy, and edge appliances hold credentials that anchor the overlay. Compromise of one management component can therefore affect many sites. MEF's SD-WAN service model captures this dependence on centralized policy and managed virtual connections [1].

The consequence is operational rather than merely architectural. In February 2026, the U.S. Cybersecurity and Infrastructure Security Agency issued Emergency Directive 26-03 after active exploitation of Cisco SD-WAN systems, followed by hunt and hardening guidance [2, 3]. The incident illustrates why published severity is not a complete remediation order. Urgency also depends on whether exploitation is observed, whether the vulnerable service is reachable now, what privilege follows from compromise, and what other assets become reachable.

CVSS describes intrinsic technical severity [4]. EPSS estimates the probability of exploitation within 30 days [5, 6]. CISA's KEV catalog records confirmed exploitation [7], while SSVC combines exploitation and mission context through decision trees [8]. These signals answer different questions. A useful queue must retain their provenance rather than treating them as interchangeable.

PECR treats prioritization as a decision over a vulnerability v, asset a, and observation time t. It produces an explained work queue from evidence already available to network and security operations. It does not deploy patches, estimate monetary loss, or replace human change control. A separate C-BOM queue handles cryptographic migration so that long-horizon transition planning is not misrepresented as evidence of an exploitable software flaw.

### 1.1 Contributions

- A reproducible, vendor-neutral specification that joins vulnerability intelligence, stable asset identity, live exposure, bounded reachability, consequence, and evidence quality at a defined observation time.
- An operational decision protocol that separates score, confidence, uncertainty bounds, exception handling, and a linked but independent cryptographic-migration queue.
- A corrected synthetic artifact with role-consistent assets, five declared baseline queues, 30,000 joint weight draws, a missing-evidence test, and 30 generator replications, accompanied by explicit limits on what those tests establish.

### 1.2 Research questions

- RQ1 — Specification: Can the evidence, factors, queue order, uncertainty rules, and cryptographic queue be defined precisely enough for independent implementation?
- RQ2 — Behavior: How does the resulting operational order differ from severity-, exploitation-, and simpler context-based heuristics on a declared synthetic cohort?
- RQ3 — Robustness and limits: How stable is the order under joint policy-weight changes and missing evidence, and which structural or adversarial conditions constrain deployment?

These questions deliberately stop short of a predictive-performance claim. Synthetic validation can test implementation behavior and robustness; production effectiveness, analyst utility, and cross-vendor transfer require independent field data.

## 2 RELATED WORK

### 2.1 Severity, exploitation, and context

Severity and exploitation are related but not interchangeable. Allodi and Massacci found that severity alone is a weak proxy for exploitation in the wild [9]. Comparisons of CVSS and EPSS support retaining a probability signal beside severity [10]. Koscinski et al. document substantial disagreement among scoring systems on 600 outcome-linked vulnerabilities [11]. Assessor experiments also show variation in structured ratings [12], and prediction is least reliable when a vulnerability has just been disclosed [13]. A recent survey consolidates the field's taxonomies and evaluation measures [14].

Composite approaches add threat signals or enterprise context. Shimizu and Hashimoto evaluate an integrated vulnerability-management chain over 28,377 real vulnerabilities [15], while Ahmadi Mehri et al. automate context-aware patch prioritization [16]. PECR follows this direction but makes live network state, bounded attack paths, and evidence quality explicit inputs. It uses EPSS probability rather than percentile because probability represents the modeled 30-day likelihood, whereas percentile changes with the scored population [5].

### 2.2 Attack paths and SD-WAN management

Attack graphs provide a basis for multi-step compromise reasoning [17, 18], but their quality remains bounded by topology, configuration, and vulnerability inputs [19]. PECR does not claim formal verification. It derives two auditable features from a bounded directed graph: distance from an untrusted origin and the share of other in-scope assets reachable within two hops.

SD-WAN surveys describe traffic engineering, service orchestration, centralized control, and security as coupled management concerns [20, 21]. Prior work has examined WAN-aware resilience [22], distributed control planes [23], and malicious-traffic detection from SD-WAN flow telemetry [24]. PECR occupies a different layer: it composes management evidence into remediation decisions rather than proposing a routing or detection mechanism.

### 2.3 Cryptographic transition and positioning

NIST standardized ML-KEM, ML-DSA, and SLH-DSA in FIPS 203–205 [25–27]. Draft NIST IR 8547 supplies a configurable transition horizon [28], CycloneDX supplies a cryptographic inventory carrier [29], Mosca's inequality links confidentiality lifetime and migration time [30], and NIST crypto-agility guidance emphasizes inventory and replaceability [31]. Federal directives add binding planning obligations for applicable systems [32, 33]. PECR uses these sources for a migration-planning queue; it does not treat planned retirement of classical cryptography as a software vulnerability.

The closest outcome-linked studies use real data at far larger scale [11, 15]. This study instead asks whether a telemetry-rich method can be specified and stress-tested when those public datasets lack the required topology and evidence-quality fields. The distinction is central: the results below describe queue behavior, not prediction accuracy.

Table 1: Positioning of PECR against representative approaches

| Approach | Threat / severity | Live context | Attack path | Evidence quality | Crypto queue |
|---|---|---|---|---|---|
| CVSS / EPSS / KEV [4–7] | Yes | No | No | No | No |
| SSVC [8] | Yes | Partial | No | No | No |
| Context-aware methods [15, 16] | Yes | Partial | No | No | No |
| Attack-graph methods [17–19] | Partial | Input dependent | Yes | No | No |
| SD-WAN management [22–24] | Partial | Yes | Partial | No | No |
| C-BOM / crypto agility [28–31] | No | Inventory | No | No | Yes |
| PECR | Yes | Yes | Bounded | Yes | Separate |

## 3 DECISION MODEL AND ARCHITECTURE

### 3.1 Unit of analysis and scope

A PECR remediation record is x = (v, a, t): a vulnerability or finding v mapped to an in-scope SD-WAN asset a at observation time t. In-scope assets include orchestrators, controllers, edge appliances, gateways, management interfaces, identity dependencies, and the routing and segmentation state required for reachability analysis. Unrelated application and endpoint findings are excluded. The defender is assumed to possess a scanner, asset inventory, threat feeds, selected flow and log sources, and a human change-control process.

### 3.2 Attack model

PECR must consider both the adversary who attacks the network and the adversary who attacks the evidence pipeline to change the queue. The latter may submit, delay, suppress, or corrupt observations from a source it controls, but cannot forge an uncompromised source credential or alter a finalized append-only decision record. The model does not assume that an authenticated observation is truthful.

Table 2: Attacks on the PECR decision layer

| ID | Attack | Mechanism | Primary effect |
|---|---|---|---|
| A1 | Telemetry poisoning | Forged or replayed source observations | Lowers or raises F4, F6, or F7 |
| A2 | Contradiction injection | Conflicting valid observations | Widens intervals; consumes analyst capacity |
| A3 | Path spoofing | Policy or deployed-state manipulation | Misstates F6 and F7 |

| ID | Attack | Mechanism | Primary effect |
|---|---|---|---|
| A4 | Freshness starvation | Feed suppression or delay | Lowers evidence quality and confidence |
| A5 | Insider score gaming | Governed metadata changed by an authorized owner | Changes F8, F9, or policy parameters |

### 3.3 Design requirements

- Traceability and time: each factor retains source identity, event and ingest time, transformation, quality, and policy version.
- Uncertainty preservation: missing and contradictory evidence produces explicit lower and upper bounds rather than silent midpoint imputation.
- Governed decisions: weights, thresholds, graph scope, and cryptographic horizons are versioned, replayed, approved, and reversible.
- Separate outputs: remediation and cryptographic migration share stable asset identity but retain distinct scores, queues, owners, and time horizons.
- Human authority and pipeline integrity: no automatic patching occurs; sources are authenticated, sequence-checked, schema-validated, freshness-monitored, and quarantined when invalid.

### 3.4 Four planes and two queues

The evidence plane ingests static and dynamic sources. The context plane resolves stable asset identity, observation time, and directed reachability. The analysis plane calculates score, confidence, and uncertainty bounds. The decision plane produces two independent queues and records analyst feedback. The remediation queue is sorted by final band, then descending score R, then stable record identifier. Evidence-limited records enter a verification workflow; an active exception floor is retained. The migration queue is sorted independently by asset urgency $G(a)$.

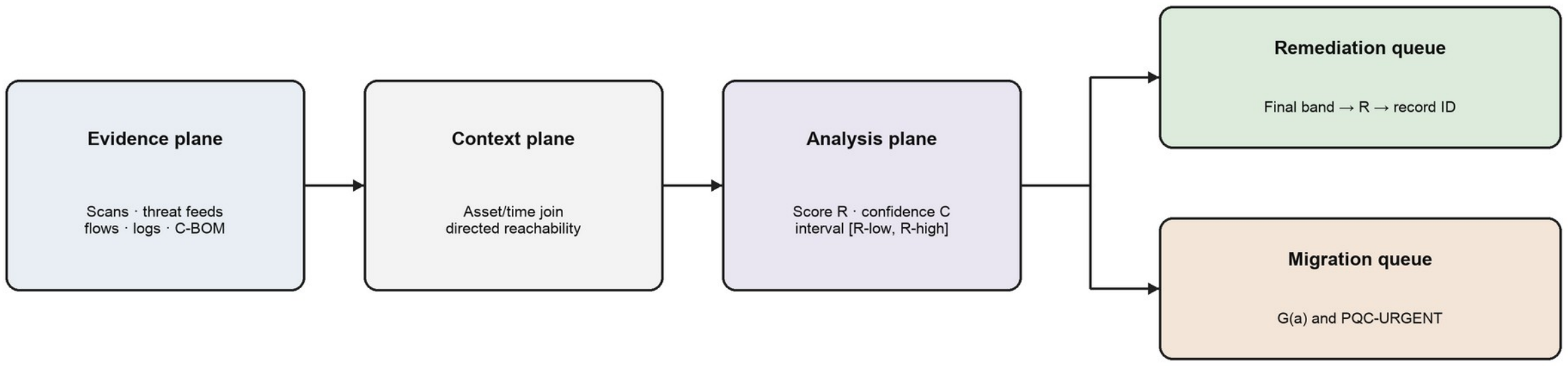

Figure 1: PECR evidence flow and separation of remediation from cryptographic migration.

# 4 EVIDENCE AND TELEMETRY MODEL

## 4.1 Source model

The minimum evidence set contains asset inventory, vulnerability findings, threat intelligence, exposure observations, identity and privilege state, routing and segmentation state, flow records, and control-plane events. IPFIX supplies a vendor-neutral flow schema [34], syslog supplies a common event envelope [35], and NTP-compatible synchronization supports temporal joins [36]. Each observation carries stable asset identity, event and ingest time, source identity, a source-signed event or sequence identifier, and schema and policy versions.

Collection cadences are governed operational parameters rather than universal constants. Threat feeds, exposure observations, and event logs are refreshed at least daily; inventory and scanner coverage are checked weekly; C-BOM records are reviewed monthly and after certificate or library changes. A KEV addition, newly observed Internet exposure, route-policy change, authentication anomaly, or ownership change triggers recomputation of affected records.

## 4.2 Directed reachability procedure

At time t, PECR builds a directed graph $G(t) = (V, E(t))$. Vertices are assets, network zones, identities, and management services. An edge $u \rightarrow z$ is admitted only when effective policy permits the transition and at least one configured or observed path supports it within the freshness window. Deny rules take precedence. A policy-only edge is retained at lower quality than a flow-corroborated edge. Divergent compiled and deployed policies create an unresolved edge rather than an assumed permit or deny.

For target a, breadth-first search begins at modeled untrusted origins. The shortest valid directed path determines h, the number of additional compromised nodes required after ingress, and $F6 = 1/(1+h)$; F6 is zero when no path is found under the modeled privilege. A bounded search then begins at a using the privilege attainable from v. If A is the in-scope asset set, F7 is the number of other assets reachable within two directed hops divided by |A|−1. The path, supporting edges, rejected alternatives, graph version, and privilege assumption are stored for audit.

### 4.3 Evidence quality and contradictions

Each factor has quality $q_i \in [0,1]$, defined as the product of freshness, completeness, and provenance controls. Freshness decays after a source-specific target and reaches zero at three times that target. Completeness measures required-field and expected-coverage ratios. Provenance is highest for authenticated, schema-valid sources and lower for inferred or operator-entered evidence.

Invalid observations are quarantined and never enter a calculation. Conflicting observations that are individually valid are not averaged: PECR retains their minimum and maximum plausible factor values, lowers $q_i$, and opens a contradiction work item. This preserves honest disagreement but creates the denial-of-attention risk analyzed in Section 7.

## 5 REMEDIATION SCORE, CONFIDENCE, AND DECISION RULES

### 5.1 Factors and score

For a complete record, PECR uses the nine normalized factors in Table 3. Displayed weights are rounded to six decimals; the executable vector is normalized to sum exactly to one.

Table 3: PECR remediation factors and AHP-derived weights

| Factor | Operational meaning | Normalization | Weight |
|---|---|---|---|
| F1 Base severity | Intrinsic technical severity | CVSS base / 10 | 0.100493 |
| F2 Exploit evidence | Observed or published exploitation | none 0; public exploit 0.5; confirmed / KEV 1 | 0.195021 |
| F3 Exploit probability | Estimated exploitation within 30 days | EPSS probability | 0.100493 |
| F4 Untrusted accessibility | Current reachability of vulnerable service | none 0; internal .33; partner .67; Internet 1 | 0.195021 |
| F5 Attainable privilege | Privilege after successful exploitation | none 0; user .33; admin .67; control plane 1 | 0.100493 |
| F6 Path reachability | Shortest modeled path from an untrusted origin | 1/(1+h); unreachable 0 | 0.100493 |
| F7 Blast radius | Other in-scope assets reachable within two hops | reachable others / (\|A\|−1) | 0.053848 |
| F8 Operational consequence | Service consequence of asset compromise | low .25; moderate .5; high .75; critical 1 | 0.105843 |
| F9 Control concentration | Authority of the asset role | application .2 to orchestrator 1 | 0.048292 |

$$R = 100 \sum_{i=1}^{9} w_i f_i, \quad \sum_{i=1}^{9} w_i = 1. \tag{1}$$

The starting vector comes from an author-elicited reciprocal AHP matrix [37]. The principal eigenvector gives $\lambda_{max} = 9.0266$ and consistency ratio 0.0023. That ratio establishes internal coherence only; it does not turn one analyst's judgments into expert consensus or an empirical optimum.

## 5.2 Confidence and interval-valued evidence

$$C = \sum_{i=1}^{9} w_i q_i, \quad 0 \le C \le 1. \tag{2}$$

Confidence is reported beside R and never multiplied into it. A high score with low confidence requires verification; a low score with low confidence is not treated as safe. To represent all evidence states consistently, each factor i has a plausible interval $[\ell_i, u_i]$. Known values use $\ell_i = u_i = f_i$, unknown values use [0,1], and contradictory valid observations use their retained minimum and maximum.

$$R^- = 100 \sum_{i=1}^{9} w_i \ell_i, \quad R^+ = 100 \sum_{i=1}^{9} w_i u_i. \tag{3}$$

A point band is issued only when $R^-$ and $R^+$ fall in the same band. Otherwise the record is labeled evidence-limited and routed to verification before normal scheduling. This interval equation, unlike a missing-only shortcut, also encodes contradictory factor ranges.

## 5.3 Bands, exception, and deterministic queue order

Default calculated bands are Critical ($R \ge 85$), High (70–84.99), Medium (50–69.99), Low (30–49.99), and Monitor ($R < 30$). They are service categories, not probabilities. Exception E1 sets the final band to at least High when F2 = 1 and F4 = 1, meaning confirmed exploitation and confirmed Internet accessibility. E1 changes neither R nor C.

The operational sort key is: descending final-band severity, descending R, then stable record identifier. The identifier is used only to make exact ties reproducible. This definition separates an operational total order from inferential rank statistics and makes clear how E1 affects the queue.

## 5.4 Elicited weight versus realized influence

Nominal weight is not realized influence. A factor's share of score mass depends on its observed distribution as well as $w_i$. In the corrected primary cohort, F2 holds 19.5% nominal weight but contributes 6.2% of score mass, while F8 holds 10.6% nominal weight but contributes 15.7%. Deployments should report both quantities and should not interpret the AHP vector as a measured causal contribution.

## 6 PARALLEL CRYPTOGRAPHIC-MIGRATION QUEUE

### 6.1 Bounded C-BOM and normalized components

The C-BOM covers cryptography that operates the SD-WAN: overlay establishment, controller and management sessions, device and administrator authentication, certificate chains, and protection of configuration and backups. Application-layer cryptography merely transiting the WAN is outside scope. Records identify asset, protected function, protocol, library and version, certificate, algorithms, key-establishment mechanism, data sensitivity, confidentiality lifetime, migration duration, dependencies, owner, status, constraints, and retirement date. CycloneDX v1.6 is a suitable carrier [29], but the fields—not the carrier—define PECR's requirement.

Table 4: Default normalization for cryptographic-migration components

| Component | Meaning | Default normalization |
|---|---|---|
| $c_1$ | Algorithm lifecycle | approved 0; deprecation announced .5; disallowed or beyond retirement 1 |
| $c_2$ | Time pressure | $\min\{1,(L+M)/H\}$ |
| $c_3$ | Exposure of protected data | offline 0; internal .33; partner .67; public 1 |
| $c_4$ | Dependency depth | 0, .33, .67, 1 for none, one, two, or ≥3 blocking layers |
| $c_5$ | Migration complexity | routine .25; moderate .5; high .75; blocked / redesign 1 |
| $c_6$ | Regulatory obligation | none 0; planning obligation .5; binding dated mandate 1 |

$$g(d) = 0.25c_1 + 0.25c_2 + 0.20c_3 + 0.10c_4 + 0.10c_5 + 0.10c_6. \qquad (4)$$

For confidentiality lifetime $L$, estimated migration duration $M$, and configured planning horizon $H$, $c_2$ reaches one when $L+M \geq H$. This means the combined confidentiality and migration requirement meets or exceeds the horizon; it does not imply $L \geq H$ or predict the arrival of a cryptographically relevant quantum computer.

$$G(a) = \max \{ g(d) : d \in D(a) \}. \qquad (5)$$

The maximum is used because one blocking dependency can determine an asset's migration schedule. PQC-URGENT is raised when $c_2 = 1$ and $c_3 \geq 0.67$: the combined time requirement reaches the planning horizon and protected data is exposed on partner or public paths. The flag and $G(a)$ enter only the migration queue.

Example: an Internet-transiting IPsec dependency with $c = (0.7, 1, 1, 0.5, 0.5, 0.5)$ has $g(d) = 0.775$ and activates PQC-URGENT. The dependency enters migration review; any software vulnerability on the same asset remains ordered by the remediation protocol.

## 7 SECURITY ANALYSIS OF THE EVIDENCE PIPELINE

Table 5 evaluates the attacks from Section 3.2. The analysis is a design argument, not a proof: controls are conditional on external evidence sources and organizational authority.

Table 5: Specified controls and residual risk

| Attack | Specified controls | Residual condition |
|---|---|---|
| A1 Telemetry poisoning | Source authentication; signed event/sequence IDs; replay window; schema validation; independent flow corroboration | A compromised authoritative source can submit plausible false observations |
| A2 Contradiction injection | Retain min/max; reduce quality; monitor contradiction rate; preserve E1 floor | A valid source can exhaust analyst capacity; per-source rate limiting remains future work |
| A3 Path spoofing | Deny precedence; compiled/deployed comparison; graph versioning; retained alternatives | Control of the policy system of record can make wrong views agree |
| A4 Freshness starvation | Heartbeat and cadence alerts; freshness decay; upper-bound review | Operators may still defer low-confidence records unless a queue floor is enforced |
| A5 Insider gaming | Versioned parameters; replay; separation of duties; append-only decision records | Legitimate authority can still be misused within role |

For A1 and A3, authentication proves source identity, not truth. Flow corroboration and retained graph alternatives can support later diagnosis if independent evidence emerges, but they do not guarantee detection of a compromised authoritative source. Append-only records provide auditability; they are not themselves a detector.

A2 and A4 expose a deeper tension. Widening intervals and lowering confidence is appropriate for honest disagreement or missing data, yet an adversary can exploit the same behavior to create analyst load. E1 prevents suppression of the confirmed-exploitation/Internet-accessible class, but it does not bound the general attack. A production deployment should rate-limit contradiction work per source and apply a queue floor when low confidence coexists with a high upper bound.

A5 is primarily organizational. Separation of duties should keep consequence classification and remediation accountability in different roles. PECR can record and replay changes, but it cannot prevent an authorized owner from exercising legitimate authority badly.

## 8 SYNTHETIC VALIDATION

### 8.1 Generator, asset consistency, and scope

The corrected artifact generates 100 vulnerability–asset records over 62 synthetic assets with fixed seed 20260731. Asset roles are fixed across records: 4 orchestrators, 6 controllers, 15 Internet edges, 19 branch edges, and 6 each of identity services, guest gateways, and core gateways. Record-role counts are 6, 10, 24, 30, 10, 10, and 10, respectively. Records are assigned round-robin within role and then shuffled; every repeated asset identifier therefore retains one role, zone, consequence value, and control-concentration value.

The generator is role- and zone-conditioned, not a packet-level topology simulator. A latent threat $z \sim \text{Beta}(1.7,3.0)$ is shared by EPSS and exploit-evidence generation. CVSS is clipped to [0.1,10] after $3 + 5.5\cdot\text{Beta}(2.2,1.7) + \text{Normal}(0,0.55)$. EPSS uses logistic($-5.1 + 0.38\cdot\text{CVSS} + 3z + \text{Normal}(0,0.65)$); KEV uses a Bernoulli probability logistic($-6 + 0.30\cdot\text{CVSS} + 3.8z + 0.65\cdot\text{IInternet}$); and non-KEV public-exploit evidence uses logistic($-4.2 + 0.27\cdot\text{CVSS} + 2.6z$). Other factors use declared role and zone profiles in the artifact. The study samples F6 and F7; it does not validate the graph-construction algorithm itself.

The artifact contains the generator, full AHP matrix, primary CSV, baseline results, joint-sensitivity summary, replication summary, and machine-readable JSON. It validates arithmetic and declared stress behavior only. It uses no production telemetry or observed exploitation outcome.

AI-use disclosure. OpenAI Codex was used during manuscript revision to audit implementation logic, propose corrections to the synthetic validation code, assist with execution and result checks, and generate revised figures and prose. Its outputs were not treated as evidence: every reported value is generated by the disclosed executable artifact, byte-level reproducibility was checked using the fixed seed, and cited claims were checked against publisher or primary records. The human authors retain responsibility for independently reviewing the artifact and for all methods, results, and claims before submission. No AI system supplied production telemetry, outcome labels, or external experimental observations.

### 8.2 Baselines and ranking protocol

Five comparators are declared before analysis: CVSS-only with EPSS tie-break; EPSS-only with CVSS tie-break; KEV-first with CVSS then EPSS tie-break; CVSS×EPSS with KEV tie-break; and context-lite. Context-lite uses 0.25F1 + 0.25F2 + 0.15F3 + 0.20F4 + 0.15F8, applies E1, and uses the same final-band/score/identifier protocol as PECR. The last comparator asks whether PECR's additional privilege, path, blast-radius, and control-concentration factors add substantial ordering change beyond a simpler context model.

Table 6: Primary cohort—queue agreement with PECR

| Queue | Kendall τ | Top-10 overlap | Top-10 KEV / Internet / control-plane |
|---|---|---|---|
| PECR | 1.000 | 10 | 7 / 9 / 1 |
| CVSS-only | 0.196 | 1 | 2 / 1 / 4 |
| EPSS-only | 0.221 | 2 | 2 / 3 / 1 |
| KEV-first | 0.301 | 7 | 8 / 6 / 2 |
| CVSS×EPSS | 0.234 | 2 | 1 / 3 / 1 |
| Context-lite | 0.755 | 8 | 8 / 7 / 1 |

Context-lite is the closest comparator ($\tau = 0.755$; top-10 overlap 8), substantially closer than KEV-first ($\tau = 0.301$; overlap 7). Thus most of the operational top-10 behavior is already captured by confirmed exploitation, Internet accessibility, severity, probability, and consequence. PECR's additional factors primarily refine within- and near-band ordering in this synthetic cohort. That is evidence of incremental differentiation, not superiority.

### 8.3 Joint global weight sensitivity

Complete weight vectors are drawn from Dirichlet(κw), a distribution on the nine-part simplex [38, 39]. All weights change jointly and continue to sum to one. Concentrations κ = 20, 50, and 100 represent broad, moderate, and narrow stress regimes, with 10,000 draws each. They are sensitivity scenarios, not a posterior distribution of stakeholder belief.

Table 7: Joint-weight sensitivity, 10,000 draws per concentration

| κ | τ p05 / med / p95 | Top-10 p05 / med / p95 | Band agreement p05 / med / p95 | Least-stable base top-10 record |
|---|---|---|---|---|
| 20 | 0.664 / 0.836 / 0.917 | 7 / 9 / 10 | 0.52 / 0.78 / 0.91 | 57.9% |
| 50 | 0.788 / 0.893 / 0.949 | 8 / 10 / 10 | 0.68 / 0.86 / 0.96 | 70.1% |
| 100 | 0.846 / 0.924 / 0.964 | 8 / 10 / 10 | 0.78 / 0.90 / 0.98 | 80.1% |

Under broad perturbation, median τ is 0.836 and median top-10 overlap is nine, but the least-stable base top-10 record is retained in only 57.9% of draws. Median stability should therefore not be interpreted as uniform record-level stability. The operational band comparison includes E1 and the declared final-band sort protocol.

### 8.4 Band calibration, missing evidence, and realized influence

The primary cohort has median R = 43.65, 95th percentile 68.19, and maximum 76.47. Calculated bands contain 18 Monitor, 51 Low, 26 Medium, 5 High, and 0 Critical records. E1 applies to six records but changes the band of only two; final counts are 18, 51, 24, 7, and 0. The unpopulated Critical band shows that the default thresholds are not calibrated to this generator. Deployments should calibrate service bands on governed historical outcomes rather than adopt these defaults unchanged.

A separate test masks 90 of 900 factor cells (10%) uniformly without replacement. Sixty-three records lose at least one factor. Equation (3) produces a median interval width of 14.88 points and a 95th-percentile width of 29.55; 47 intervals cross a calculated-band boundary. The test demonstrates interval propagation under one missing-at-random mechanism. It does not show that the labels are correct against ground truth or model adversarial missingness.

Realized score mass also differs from nominal policy weight. Exploit evidence and exploit probability contribute less than their weights in this cohort because they are often near zero; operational consequence, accessibility, privilege, and severity contribute more. This finding argues for reporting realized influence beside elicited weight during deployment review.

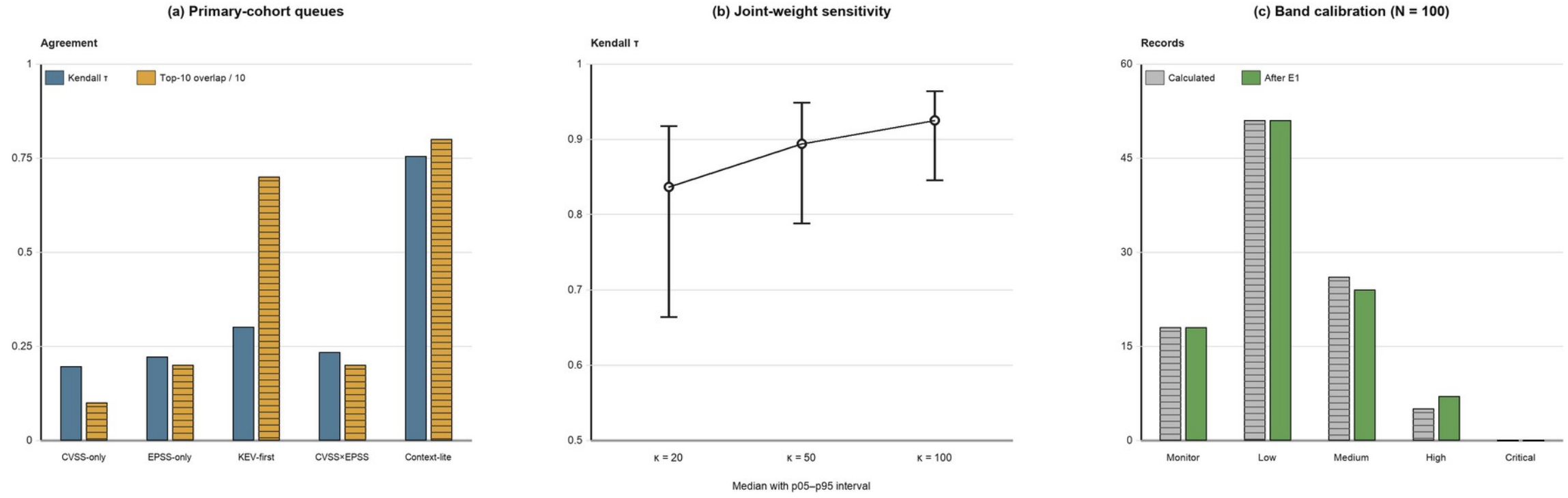

Figure 2: Corrected validation summary for queue agreement, joint-weight sensitivity, and band calibration.

### 8.5 Generator replication

The generator is rerun 30 times with 100 records per seed, adding 3,000 records. Replication reduces dependence on one seed but retains the same data-generating assumptions.

Table 8: Thirty-seed generator replication

| Queue | τ p05 / median / p95 | Top-10 overlap p05 / median / p95 |
|---|---|---|
| CVSS-only | 0.069 / 0.153 / 0.276 | 0.45 / 2.00 / 4.00 |
| EPSS-only | 0.079 / 0.173 / 0.270 | 1.00 / 3.00 / 4.00 |
| KEV-first | 0.172 / 0.266 / 0.366 | 3.00 / 6.00 / 8.00 |
| CVSS×EPSS | 0.090 / 0.180 / 0.280 | 0.45 / 2.00 / 4.00 |
| Context-lite | 0.686 / 0.727 / 0.777 | 6.00 / 8.00 / 9.00 |

Context-lite remains the closest comparator across seeds (median τ = 0.727; median top-10 overlap 8). KEV-first remains next closest (median τ = 0.266; median overlap 6). These results support a qualitative distinction between adding basic context and adding PECR's full factor set, while remaining conditional on one synthetic generator family.

### 8.6 What the validation establishes

The corrected study establishes executable arithmetic, deterministic queue semantics, differentiation from five declared heuristics, robustness under stated weight distributions, propagation of missing inputs, and repeatability across seeds. It does not establish predictive accuracy, prevented exploitation, avoided loss, analyst acceptance, or a universally appropriate weight vector. Those outcomes require independent retrospective and prospective field studies.

## 9 ADOPTION AND GOVERNANCE

### 9.1 Staged adoption

PECR can be introduced incrementally. Level 0 establishes stable asset identity, vulnerability coverage, and consequence ownership. Level 1 adds threat intelligence and exposure observations. Level 2 adds continuous telemetry and evidence quality. Level 3 activates directed paths and blast radius. Level 4 populates the bounded C-BOM. Level 5 introduces retrospective outcome review and governed calibration.

Initial readiness thresholds may include 90% asset-role and consequence coverage at Level 0; 95% feed freshness at Level 1; median confidence of at least 0.70 at Level 2; at least 80% sampled path validation at Level 3; and C-BOM coverage of all orchestrators and controllers plus half of edges at Level 4. These are starting controls, not validated universal benchmarks.

### 9.2 Privacy, change control, and audit

PECR requires operational metadata but not packet payloads. Flow records are minimized to reachability fields, identity values are represented by stable pseudonymous identifiers outside the identity system, and analyst access is logged. Raw telemetry follows source-system retention;

PECR retains factor values, provenance, policy version, and append-only decision records for replay.

Weights, thresholds, normalizations, graph scope, service levels, and cryptographic horizons are governed configuration. Each change requires rationale, owner, replay against an archived evaluation set, comparison of rank and band changes, approval, effective date, monitoring period, and rollback plan. PECR does not modify production policy autonomously. This governance is consistent with the govern, identify, and improve functions of NIST CSF 2.0 [40].

## 10 LIMITATIONS AND VALIDATION AGENDA

The nine factors are plausible operational constructs, but their ordinal scales remain policy choices. AHP makes one starting vector traceable; it does not make one analyst representative of network operators. F6 and F7 depend on topology correctness, and F7 can overlap with F9 in centralized environments. EPSS is a model output rather than a guarantee. The C-BOM horizon depends on evolving guidance and local migration estimates.

Internal validity is limited by role counts, conditional distributions, and the shared latent-threat construction. The corrected generator enforces asset-level consistency, but it still samples reachability and blast radius instead of executing the graph procedure. Thirty replications test seeds, not alternative models. Uniform masking is only one missingness mechanism. The cryptographic queue is specified and illustrated but not empirically evaluated.

External validity is the principal limitation. The 100 primary and 3,000 replication records do not represent vendor differences, maintenance windows, patch dependencies, attacker adaptation, or organizational incentives. Kendall τ and top-k overlap measure queue disagreement, not accuracy or significance. Confidence measures evidence accounting, not a calibrated probability of correctness.

The next evaluation should preregister a retrospective, multi-environment study. Baselines should include CVSS, EPSS, KEV-first, SSVC, and context-lite. Outcomes should include precision at fixed remediation budgets, time to mitigate later-exploited findings, analyst top-k agreement, path-validation rate, queue churn, and confidence calibration. A prospective study should then measure decision time and override quality without permitting autonomous deployment. Stakeholder AHP panels should report inter-rater dispersion and realized influence.

## 11 CONCLUSION

PECR is a reproducible decision specification for ordering SD-WAN remediation while keeping evidence quality visible. The method separates risk score, confidence, plausible bounds, policy exceptions, and cryptographic migration. The corrected artifact clarifies asset identity, operational queue order, interval mathematics, cryptographic normalization, and the difference between simple and full context.

The strongest result is appropriately modest: basic context accounts for much of PECR's top-10 behavior, while the full factor set refines the order and remains reasonably stable under the declared weight stresses. The evidence supports synthetic feasibility and exposes important calibration and pipeline-security limits. It does not support a claim of production effectiveness; independent field validation remains the necessary next step.

## 12 CODE AND DATA AVAILABILITY

A supplementary artifact accompanies this paper. It contains the corrected generator, fixed seeds, synthetic primary dataset, AHP matrix, baseline results, sensitivity and replication summaries, and machine-readable JSON. A permanent archival identifier will be added in a subsequent version.